\def\half {\mbox{$\textstyle {1 \over 2}$}}

\documentclass[12pt,a4paper]{article}
\usepackage{amsmath,amssymb,epsfig,array,lscape}

\renewcommand{\title}[1]{{\Large\bf\mbox{}\\\medskip#1\bigskip\medskip\\ }}
\renewcommand{\author}[1]{{\large #1\smallskip\\ }}
\newcommand{\address}[1]{{\em #1\medskip\\ }}
\def\Z{{\mathbb Z}}

\def\hbar{\overline{h}}
\def\Tr{\mbox{Tr}}
\def\Re{\mbox{Re}}
\def\bra#1{\langle #1|}
\def\ket#1{| #1\rangle}
\def\V{{\cal V}}
\def\H{{\cal H}}
\def\F{{\cal F}}
\def\B{{\cal B}}
\def\P{{\cal P}}
\def\M{{\cal M}}
\def\vm{\boldsymbol{m}}
\def\vn{\boldsymbol{n}}
\def\ve{\boldsymbol{e}}
\def\D{\boldsymbol{D}}
\def\Q{\boldsymbol{Q}}

\def\gauss#1#2{\mbox{\small $\left[#1\atop #2\right]$}}

\newcommand{\sm}[1]{{\scriptstyle #1}}

\newcommand{\spos}[2]{\makebox(0,0)[#1]{$\sm{#2}$}}
\newcommand{\sposb}[2]{\makebox(0,0)[#1]{$ #2 $}}

\newcommand{\W}[5]{W\!\left(\,\begin{array}{@{}cc|@{\:}}#4&#3\\
#1&#2\end{array}\;#5\right)}
\newcommand{\face}[5]{\begin{picture}(1.6,1.6)
\multiput(0.3,0.3)(1,0){2}{\line(0,1){1}}
\multiput(0.3,0.3)(0,1){2}{\line(1,0){1}}
\put(0.26,0.26){\spos{tr}{#1}}\put(1.34,0.26){\spos{tl}{#2}}
\put(1.34,1.34){\spos{bl}{#3}}
\put(0.26,1.34){\spos{br}{#4}}
\put(0.8,0.8){\spos{}{#5}}\end{picture}}

\def\lattices{\small{\put(-5,0){\makebox(0,0)[r]{$1$}}
\put(-5,20){\makebox(0,0)[r]{$2$}}
\put(-5,40){\makebox(0,0)[r]{$3$}}
\put(-5,60){\makebox(0,0)[r]{$4$}}
\put(0,-5){\makebox(0,0)[t]{$0$}}
\put(20,-5){\makebox(0,0)[t]{$1$}}
\put(40,-5){\makebox(0,0)[t]{$2$}}
\put(60,-5){\makebox(0,0)[t]{$3$}}
\put(80,-5){\makebox(0,0)[t]{$4$}}
\put(100,-5){\makebox(0,0)[t]{$5$}}
\put(120,-5){\makebox(0,0)[t]{$6$}}
\put(140,-5){\makebox(0,0)[t]{$7$}}
\put(160,-5){\makebox(0,0)[t]{$8$}}
\multiput(0,0)(0,20){4}{\line(1,0){160}}
\multiput(0,0)(20,0){9}{\line(0,1){60}}}}

\def\latticel{\small{\put(-5,0){\makebox(0,0)[r]{$1$}}
\put(-5,20){\makebox(0,0)[r]{$2$}}
\put(-5,40){\makebox(0,0)[r]{$3$}}
\put(-5,60){\makebox(0,0)[r]{$4$}}
\put(0,-5){\makebox(0,0)[t]{$0$}}
\put(20,-5){\makebox(0,0)[t]{$1$}}
\put(40,-5){\makebox(0,0)[t]{$2$}}
\put(60,-5){\makebox(0,0)[t]{$3$}}
\put(80,-5){\makebox(0,0)[t]{$4$}}
\put(100,-5){\makebox(0,0)[t]{$5$}}
\put(120,-5){\makebox(0,0)[t]{$6$}}
\put(140,-5){\makebox(0,0)[t]{$7$}}
\put(160,-5){\makebox(0,0)[t]{$8$}}
\put(180,-5){\makebox(0,0)[t]{$9$}}
\put(200,-5){\makebox(0,0)[t]{$10$}}
\put(220,-5){\makebox(0,0)[t]{$11$}}
\put(240,-5){\makebox(0,0)[t]{$12$}}
\multiput(0,0)(0,20){4}{\line(1,0){240}}
\multiput(0,0)(20,0){13}{\line(0,1){60}}}}

\begin{document}
\hfill\today
\begin{center}
\title{Critical RSOS and Minimal Models I: \\
Paths, Fermionic Algebras and Virasoro Modules}
\author{ Giovanni Feverati\footnote{
Email: feverati@ms.unimelb.edu.au} and Paul A. Pearce\footnote{
Email: P.Pearce@ms.unimelb.edu.au}}
\address{Department of Mathematics and Statistics\\
University of Melbourne\\Parkville, Victoria 3010, Australia}
\end{center}
\begin{abstract}
We consider $s\ell(2)$ minimal conformal field theories on a cylinder from a lattice perspective. To each allowed one-dimensional configuration path of the $A_L$ Restricted Solid-on-Solid (RSOS) models  we associate a physical state and a monomial in a finite fermionic algebra. The orthonormal states produced by the action of these monomials on the primary states  $|h\rangle$ generate finite Virasoro modules with dimensions given by the finitized Virasoro characters $\chi^{(N)}_h(q)$. These finitized characters are the generating functions for the double row transfer matrix spectra of the critical RSOS models. We argue that a general energy-preserving bijection exists between the one-dimensional configuration paths and the eigenstates of these transfer matrices and exhibit this bijection for the critical and tricritical Ising models in the vacuum sector. 
Our results extend to ${\mathbb Z}_{L-1}$ parafermion models by duality.
\end{abstract}

\section{Introduction}
\setcounter{equation}{0}

Over the last two decades there has been much progress in understanding the deep connections between conformal field theory (CFT)~\cite{BPZ,FMS} and integrable lattice models~\cite{BaxBook}. But there remain some intriguing mysteries. For example it is well-known, from the work of the Stony Brook group, that the Virasoro characters admit a fermionic quasi-particle  interpretation~\cite{McCoyEtAl} and this has led to many recent developments (see for example \cite{spinons} and \cite{HKOTT} and references therein). But what is the fermionic algebra underlying the structure of the finitized characters and what are the fermionic quasi-particles? There is also a well-known correspondence principle~\cite{Kyoto} which states that the Corner Transfer Matrix (CTM) one-dimensional configurational sums of the RSOS  lattice models~\cite{ABF} in the off-critical Regimes III and II give rise to the finitized characters of the $s\ell(2)$ unitary minimal and ${\mathbb Z}_k$ parafermion models respectively. But off-criticality there is no conformal symmetry or Virasoro algebra. So how do these one-dimensional configurational sums appear in the treatment of the critical RSOS models where the conformal algebra really is present? On the other hand, workers in the field have been trying for a long time, to build~\cite{RosgenV} matrix representations of the Virasoro algebra based on paths and to make sense~\cite{ItoyamaThacker,KooS} of a finitized Virasoro algebra. We address these questions in this series of papers building on work~\cite{Warnaar} on a fermionic interpretation of Baxter's Corner Transfer Matrix (CTM) paths.

\newpage
The layout of this paper is as follows. In Section~2, we define the critical RSOS models and their double-row transfer matrices. We also recall their relation to the minimal and parafermion CFTs. 
In Section~3, we introduce fermionic algebras and paths. We give the relation between physical states and fermionic paths and discuss fermionic states and finite Virasoro modules.  
The bijection between the one-dimensional RSOS configuration paths and eigenstates of the critical RSOS double row transfer matrices is exhibited for the critical and tricritical Ising models in Section~4. 
We conclude in Section~5 with a discussion of open questions for further research.

\section{Critical RSOS Models and Conformal Field Theory}
\setcounter{equation}{0}

\subsection{Critical RSOS models and double row transfer matrices}

The critical $A_L$ RSOS models~\cite{ABF} are exactly solvable models~\cite{BaxBook} on the square lattice. 
The Boltzmann weights associated to the elementary faces are
\setlength{\unitlength}{10mm}
\begin{equation}
\W{a}{b}{c}{d}{u}=
\raisebox{-0.7\unitlength}[0.8\unitlength][0.7\unitlength]
{\begin{picture}(1.14,1.2)
\put(0.57,0.8){\setlength{\unitlength}{0.71\unitlength}
\makebox(0,0){\face{a}{b}{c}{d}{u}}}\end{picture}}
\ =\ {\sin(\lambda-u)\over \sin\lambda}\,\delta_{a,c}+
{\sin u\over\sin\lambda}\sqrt{S_aS_c\over S_bS_d}\,\delta_{b,d}
\end{equation}
where the heights $a,b,c,d\in\{1,2,\ldots,L\}$ and $u$ is the spectral parameter. The crossing parameter is $\lambda={(p-p')\pi\over p}$ with $p,p'$ coprime and $p'<p=L+1$. The crossing factors are $S_a=\sin a\lambda$ and the weights vanish if the heights on any edge do not differ by $\pm 1$. For {\em unitary} models ($p'=p-1$), these Boltzmann weights are all nonnegative whereas, for {\em nonunitary} models ($p'<p-1$), some of these weights are negative. 

Since the RSOS face weights satisfy the Yang-Baxter equation (YBE), these models are integrable for arbitrary complex $u$ using commuting transfer matrix methods~\cite{BaxBook}. To work on a strip or cylinder, with specified boundary conditions $(r_L,s_L)$ on the left and $(r_R,s_R)$ on the right edges, it is necessary to introduce~\cite{BPO} double row transfer matrices. In this paper we will consider just the subset of boundary conditions with $(r_L,s_L)=(r,1)$ and $(r_R,s_R)=(1,s)$. We call these $(r,s)$ type boundary conditions. With $(r,s)$ boundary conditions, the double row transfer matrices are represented  diagrammatically by 
\setlength{\unitlength}{14mm}
\begin{equation}
\D(u)_{\sigma,\sigma'}
=\sum_{\tau_{0},\dots,\tau_{N}}
\raisebox{-1.4\unitlength}[1.3\unitlength][1.1\unitlength]{\begin{picture}(6.4,2.4)(0.4,0.1)
\multiput(0.5,0.5)(6,0){2}{\line(0,1){2}}
\multiput(1,0.5)(1,0){3}{\line(0,1){2}}
\multiput(5,0.5)(1,0){2}{\line(0,1){2}}
\multiput(1,0.5)(0,1){3}{\line(1,0){5}}
\put(1,1.5){\line(-1,2){0.5}}\put(1,1.5){\line(-1,-2){0.5}}
\put(6,1.5){\line(1,2){0.5}}\put(6,1.5){\line(1,-2){0.5}}
\put(0.5,0.45){\spos{t}{r}}\put(1,0.45){\spos{t}{r}}
\put(2,0.45){\spos{t}{\sigma_1}}\put(3,0.45){\spos{t}{\sigma_2}}
\put(5,0.45){\spos{t}{\sigma_{N-1}}}\put(6,0.45){\spos{t}{s}}
\put(6.5,0.45){\spos{t}{s}}
\put(0.5,2.6){\spos{b}{r}}\put(1,2.6){\spos{b}{r}}
\put(2,2.6){\spos{b}{\sigma'_{1}}}\put(3,2.6){\spos{b}{\sigma'_2}}
\put(5,2.6){\spos{b}{\sigma'_{N-1}}}\put(6,2.6){\spos{b}{s}}
\put(6.5,2.6){\spos{b}{s}}
\put(1.05,1.45){\spos{tl}{\tau_0}}\put(2.05,1.45){\spos{tl}{\tau_1}}
\put(3.05,1.45){\spos{tl}{\tau_2}}\put(4.99,1.45){\spos{tr}{\tau_{N-1}}}
\put(5.99,1.45){\spos{tr}{\tau_{N}}}
\multiput(1.5,1)(1,0){2}{\sposb{}{u}}\put(5.5,1){\sposb{}{u}}
\multiput(1.5,2)(1,0){2}{\sposb{}{\lambda\!-\!u}}
\put(5.5,2){\sposb{}{\lambda\!-\!u}}
\put(0.71,1.5){\sposb{}{\lambda\!-\!u\ \ \ }}\put(6.29,1.5){\sposb{}{u}}
\multiput(0.5,0.5)(0,2){2}{\makebox(0.5,0){\dotfill}}
\multiput(6,0.5)(0,2){2}{\makebox(0.5,0){\dotfill}}
\end{picture}}
\label{RTMdef}
\end{equation}
For integrability, the triangle boundary weights on the left and right must satisfy the boundary Yang-Baxter equation (BYBE). For each conformal boundary condition $(r_L,s_L)$ or $(r_R,s_R)$ on an edge, there is a corresponding integrable boundary condition~\cite{BPMP} given by a set of triangle boundary weights satisfying the BYBE. For $(r,s)$ type boundary conditions, the partition functions satisfy $Z_{(1,1)|(r,s)}=Z_{(r,1)|(1,s)}$. In the vacuum sector $(r_L,s_L)=(r_R,s_R)=(1,1)$, the triangle boundary weights vanish unless $\tau_0=\tau_N=2$ so the triangle boundary weights can simply be removed leaving the heights alternating between $1$ and $2$ on the left and right edges of the strip.

For boundary conditions of $(r,s)$ type, the double row transfer matrices $\D(u)$ form a commuting family $[\D(u),\D(v)]=0$. 
Consequently, they can be simultaneously diagonalized by the orthogonal matrix of eigenstates which  are independent of $u$. For a suitable choice of parameters, the double row transfer matrices are real symmetric and positive definite. 
The actual form of the eigenstates changes under an orthogonal change of basis. 
Nevertheless, these eigenstates are characterised by their associated eigenvalues $D(u)$ which are {\em independent} of the choice of basis. These eigenvalues in turn can be studied analytically by Yang-Baxter techniques and are classified according to their patterns of zeros in the complex $u$-plane. Consequently, we can use the patterns of zeros to label the eigenstates.

The $A_L$ RSOS models exhibit two distinct physical regimes. If $0<u<\lambda$, the continuum scaling limit realizes the $s\ell(2)$ minimal models. Otherwise, if $\lambda-\pi/2<u<0$, the continuum scaling limit realizes the $\Z_{L-1}$ parafermions. The conformal data is obtained from the finite-size corrections to the eigenvalues of the double row transfer matrices. In making contact with CFT, the spectral parameter is usually specialized to its {\em isotropic} value, $u=\lambda/2$ for minimal models and $u=-\lambda$ for $\Z_{L-1}$ parafermions.

\subsection{Minimal models and $\Z_k$ parafermions}
The $s\ell(2)$ minimal models~\cite{BPZ} $\mathcal{M}(p',p)$ with $p,p'$ coprime have central charges
\begin{equation}\label{centralcharge}
c= 1- \frac{6 \, (p-p')^2}{p \, p'}
\end{equation}
The conformal weights are
\begin{equation}
h=h_{r,s}={(rp-sp')^2-(p-p')^2\over 4pp'},\qquad r=1,2,\ldots,p'-1;\qquad s=1,2,\ldots,p-1
\end{equation}
and the Virasoro characters are
\begin{equation}
\chi_h(q)={q^{-c/24+h}\over (q)_\infty}
\sum_{k=-\infty}^\infty \big(q^{k(kpp'+rp-sp')}-q^{(kp+s)(kp'+r)}\big)
\label{bosonchar}
\end{equation}
where
\begin{equation}
(q)_n=\prod_{k=1}^n (1-q^k)
\label{qfactorial}
\end{equation}
The minimal models are unitary if $p-p'=\pm 1$.
We consider only the diagonal $A$-type series with $p'<p$ and use the critical Ising ${\cal M}(3,4)$, tricritical Ising ${\cal M}(4,5)$ and Yang-Lee theories ${\cal M}(2,5)$ as prototypical examples.

The  $s\ell(2)$ $\Z_k$ parafermion models~\cite{ZamFat} have central charges
\begin{equation}
c={2(k-1)\over k+2},\qquad k=2,3,\ldots
\end{equation}
We consider only the diagonal $A$-type series and we use the $\Z_3$ or hard hexagon model~\cite{BaxHH,BaxBook} as the prototypical example. The hard hexagon model is in the universality class of the 3-state Potts model so we refer to this as the 3-state Potts CFT. Generally, the characters of the $\Z_k$ models are string functions but, for the $\Z_3$ model, these are easily related to the Virasoro characters of the $\M(5,6)$ model.

The minimal and $\Z_k$ parafermion models are rational and admit a finite number of primary fields $\phi(z)=\phi^{(h)}(z)$. These theories arise from the continuum scaling limit of the $A_L$ RSOS models with $L=p-1$ and $L=k+1$ respectively.

\subsection{Virasoro algebra and Virasoro states}

The Virasoro algebra
\begin{equation}
\mbox{Vir}=\langle L_n,n\in\Z\rangle 
\end{equation}
 is an infinite dimensional complex Lie algebra associated with conformal symmetry. The generators $L_n$ satisfy the commutation relations
\begin{equation}
[L_n,L_m]=(n-m)L_{n+m}+{c\over 12}\,n(n^2-1)\delta_{n,-m}
\end{equation}
where the central element $c$ is the central charge.  
The Virasoro generators are the modes of the energy-momentum tensor
\begin{equation} \label{generators}
T(z)=\sum_{n\in\Z} L_n\, z^{-n-2}
\end{equation}
On a cylinder with prescribed boundary conditions, which is the case of primary concern here, there is just one copy of the Virasoro algebra. For bulk theories on the torus, however, there is a second copy $\overline{\mbox{Vir}}$ of Virasoro which is the antiholomorphic counterpart.

For rational CFTs, the Hilbert space ${\cal H}$ of states on which Vir acts is naturally decomposed into a finite direct sum  of  irreducible highest weight representations (Virasoro modules)
\begin{equation} \label{hilbert}
{\cal H}= \oplus_{h} \V_h
\end{equation}
where the sum is over the conformal weights $h$ of the primary fields $\phi(z)=\phi^{(h)}(z)$. 
The vacuum $\ket 0$ and primary (highest weight) states $\ket h$ are characterised by
\begin{equation}
L_0\ket h=h\ket h;\qquad L_n\ket 0=0,\ \ n\ge -1;\qquad L_n\ket h=0,\ \ n>0
\end{equation}
Moreover, there is a one-to-one correspondence between primary fields and primary states induced by 
\begin{equation} \label{prim.states}
\lim_{z\to 0} \phi^{(h)}(z)\ket 0 = \ket h
\end{equation}
The vacuum state $\ket 0$ with $h=0$ corresponds to the identity operator.

The generically reducible highest weight representation of Vir (Verma module) is the linear span of Virasoro states in the canonical form
\begin{equation} \label{linear_span}
L_{-n_j}L_{-n_{j-1}} \ldots L_{-n_1}\ket h,\qquad n_j\ge n_{j-1}\ge \cdots \ge n_1\ge 1
\end{equation}
If its maximal proper submodule is quotiented out, we are led to the irreducible Virasoro module  
$\V_h=\V_{c,h}$ and the states  (\ref{linear_span}) are no longer linearly independent due to the existence of null vectors.
The generic Virasoro module in the $h=0$ vacuum sector is shown in Figure~1. Typically, for given $c$ and $h$, some states at a given level enter in a vanishing non-trivial linear combination that is the null vector. 
Surprisingly, it seems that a complete set of linearly independent Virasoro states is not known even for the Ising model, although it is known~\cite {FNO} for the  Yang-Lee theory $\M(2,5)$ and the whole family $\M(2,2n+3)$, $n\ge 1$. 

With reference to the vectors (\ref{linear_span}), the module $\V_h$ is graded according to the level
\begin{equation}
\V_h=
\mathop{\oplus}_{\ell=0}^\infty \V_{h,\ell},\qquad \ell=\sum_{i=1}^j n_i
\end{equation}
The Virasoro character $\chi_h(q)$, which is the generating function for the spectrum of the Virasoro module $\V_h$, is
\begin{equation} \label{main_character}
\chi_h(q)= \Tr_{\V_h} q^{L_0-c/24}=q^{-c/24+h}\sum_{\ell=0}^\infty d_\ell\, q^\ell,\qquad d_\ell\ge 0
\end{equation}
where $q$ is the modular parameter and the degeneracy $d_\ell=d^h_\ell=\mbox{dim}\,\V_{h,\ell}$ is the dimension of the space of states at level $\ell$.

\setlength{\unitlength}{1pt}
\begin{figure}[htb]
\hspace{.5in}
\begin{picture}(320,240)
\put(0,220){$\phantom{2}1$}\put(40,220){$\ket 0$}
\put(0,200){$\phantom{2}0$}\put(40,200){--}
\put(0,180){$\phantom{2}q^2$}\put(40,180){$L_{-2}\ket 0$}
\put(0,160){$\phantom{2}q^3$}
\put(40,160){$L_{-3}\ket 0$}
\put(0,140){$2q^4$}
\put(40,140){$L_{-4}\ket 0$}
\put(120,140){$L_{-2}^2\ket 0$}
\put(0,120){$2q^5$}
\put(40,120){$L_{-5}\ket 0$}
\put(120,120){$L_{-3}L_{-2}\ket 0$}
\put(0,100){$4q^6$}
\put(40,100){$L_{-6}\ket 0$}
\put(120,100){$L_{-4}L_{-2}\ket 0$}
\put(200,100){$L_{-3}^2\ket 0$}
\put(280,100){$L_{-2}^3\ket 0$}
\put(0,80){$4q^7$}
\put(40,80){$L_{-7}\ket 0$}
\put(120,80){$L_{-5}L_{-2}\ket 0$}
\put(200,80){$L_{-4}L_{-3}\ket 0$}
\put(280,80){$L_{-3}L_{-2}^2\ket 0$}
\put(0,60){$7q^8$}
\put(40,60){$L_{-8}\ket 0$}
\put(120,60){$L_{-6}L_{-2}\ket 0$}
\put(200,60){$L_{-5}L_{-3}\ket 0$}
\put(280,60){$L_{-4}^2\ket 0$}
\put(60,40){$L_{-4}L_{-2}^2\ket 0$}
\put(160,40){$L_{-3}^2L_{-2}\ket 0$}
\put(260,40){$L_{-2}^4\ket 0$}
\put(0,20){$8q^9$}
\put(50,20){$\ldots$}
\put(140,20){$\ldots$}
\put(220,20){$\ldots$}
\put(300,20){$\ldots$}
\put(0,0){$12q^{10}$}
\put(50,0){$\ldots$}
\put(140,0){$\ldots$}
\put(220,0){$\ldots$}
\put(300,0){$\ldots$}
\end{picture}
\caption{Virasoro module $\V_0$ of Virasoro states in the vacuum $h=0$ sector. The generic Virasoro character is $\chi_0(q)=\prod_{n=2}^\infty (1-q^n)^{-1}
=1+q^2+q^3+2q^4+2q^5+4q^6+4q^7+7q^8+8q^9+12q^{10}+\cdots$. In this sector, there is a null vector $L_{-1}\ket 0=0$ at level  $\ell=1$. For the minimal theories $\M(p',p)$, further null vectors appear. For example, for the Ising model $\M(3,4)$, there is one null vector at level $6$ and $7$ and two at level $8$. For $\M(4,5)$, the first null vector enters at  level $12$.}
\end{figure}

\section{Fermionic Algebras and States}
\setcounter{equation}{0}

\subsection{Fermionic algebras}
\label{subsec_ferm_alg}
Consider the $N$-step configuration paths $\{\sigma\}$ of the $A_L$ RSOS models~\cite{ABF} as shown in Figures~2 and 3 with $\sigma_j\in \{1,2,3,\ldots,L\}$ and $\sigma_{j+1}-\sigma_j=\pm 1$. In this context, applying conformal boundary conditions of $(r,s)$ type means that $\sigma_0=s$, $\sigma_N=r$ and $\sigma_{N+1}=r+1$ where $s=1,2,\ldots, L$ and $r=1, 2, \ldots, L-1$. Alternatively, we can work with infinite paths which start at $s$ and after $N$ steps alternate between heights $r$ and $r+1$. 
Allowing for the ${\Z}_2$ height reversal symmetry, there are $\half L(L-1)$ distinct boundary conditions or sectors.    In the case of the unitary minimal models ${\cal M}(L,L+1)$, these are in  one-to-one correspondence with the primary operators $\phi^{(h)}(z)$ with conformal weights in the Kac table
\begin{equation}
h=h_{r,s}=h_{L-r,L+1-s}={\big((L+1)r-Ls\big)^2-1\over 4L(L+1)},\qquad 1\le r\le L-1,\quad 1\le s\le L
\end{equation}

\begin{figure}[htb]
\begin{center}
\begin{picture}(160,80)
\put(-24,0){\makebox(0,0)[r]{$s=1$}}
\put(205,0){\makebox(0,0)[r]{$r=1$}}
\put(-5,0){\makebox(0,0)[r]{$1$}}
\put(-5,20){\makebox(0,0)[r]{$2$}}
\put(-5,40){\makebox(0,0)[r]{$3$}}
\put(-5,60){\makebox(0,0)[r]{$4$}}
\put(0,-5){\makebox(0,0)[t]{$0$}}
\put(20,-5){\makebox(0,0)[t]{$1$}}
\put(40,-5){\makebox(0,0)[t]{$2$}}
\put(60,-5){\makebox(0,0)[t]{$3$}}
\put(80,-5){\makebox(0,0)[t]{$4$}}
\put(100,-5){\makebox(0,0)[t]{$5$}}
\put(120,-5){\makebox(0,0)[t]{$6$}}
\put(140,-5){\makebox(0,0)[t]{$7$}}
\put(160,-5){\makebox(0,0)[t]{$8$}}
\multiput(0,0)(0,20){4}{\line(1,0){160}}
\multiput(0,0)(20,0){9}{\line(0,1){60}}
\thicklines
\put(0,0){\line(1,1){20}}
\put(20,20){\line(1,1){20}}
\put(40,40){\line(1,1){20}}
\put(60,60){\line(1,-1){20}}
\put(80,40){\line(1,-1){20}}
\put(100,20){\line(1,1){20}}
\put(120,40){\line(1,-1){20}}
\put(140,20){\line(1,-1){20}}
\put(160,0){\line(1,1){20}}
\multiput(20,20)(4,-4){5}{.}
\multiput(60,20)(4,-4){5}{.}
\multiput(100,20)(4,-4){5}{.}
\multiput(40,0)(4,4){5}{.}
\multiput(80,0)(4,4){5}{.}
\multiput(120,0)(4,4){5}{.}
\end{picture}
\end{center}
\caption{The $N=8$ step RSOS configurational path $\sigma=\{1,2,3,4,3,2,3,2,1\}$ in the vacuum $(r,s)=(1,1)$ sector of the tricritical Ising model ${\cal M}(4,5)$ corresponding to the fermionic state  $b_{-{7\over 2}}b_{-2}b_{-1}b_{-{1\over 2}}\ket 0$. The energy of this path is $E(\sigma)=\half(1+2+4+7)=7$. The groundstate vacuum path $\ket 0$ is shown dotted.}
\end{figure}
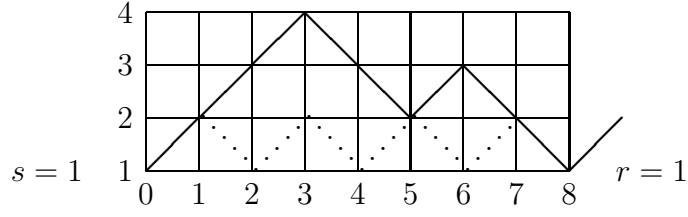
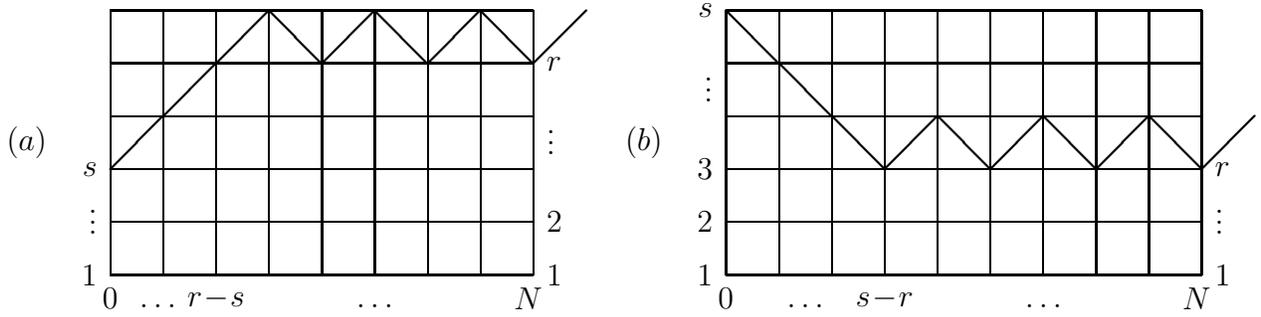
\begin{figure}[htb]
\begin{center}
\begin{picture}(160,120)
\put(-24,50){\makebox(0,0)[r]{$(a)$}}
\put(-5,0){\makebox(0,0)[r]{$1$}}
\put(-5,23){\makebox(0,0)[r]{$\vdots$}}
\put(-5,40){\makebox(0,0)[r]{$s$}}
\put(165,0){\makebox(0,0)[l]{$1$}}
\put(165,20){\makebox(0,0)[l]{$2$}}
\put(165,53){\makebox(0,0)[l]{$\vdots$}}
\put(165,80){\makebox(0,0)[l]{$r$}}
\put(0,-5){\makebox(0,0)[t]{$0$}}
\put(18,-11){\makebox(0,0)[t]{$\ldots$}}
\put(40,-5){\makebox(0,0)[t]{$r\!-\!s$}}
\put(100,-11){\makebox(0,0)[t]{$\ldots$}}
\put(158,-5){\makebox(0,0)[t]{$N$}}
\multiput(0,0)(0,20){6}{\line(1,0){160}}
\multiput(0,0)(20,0){9}{\line(0,1){100}}
\thicklines
\put(0,40){\line(1,1){20}}
\put(20,60){\line(1,1){20}}
\put(40,80){\line(1,1){20}}
\put(60,100){\line(1,-1){20}}
\put(80,80){\line(1,1){20}}
\put(100,100){\line(1,-1){20}}
\put(120,80){\line(1,1){20}}
\put(140,100){\line(1,-1){20}}
\put(160,80){\line(1,1){20}}
\end{picture}
\hspace{.9in}
\begin{picture}(160,120)
\put(-24,50){\makebox(0,0)[r]{$(b)$}}
\put(-5,0){\makebox(0,0)[r]{$1$}}
\put(-5,20){\makebox(0,0)[r]{$2$}}
\put(-5,40){\makebox(0,0)[r]{$3$}}
\put(-5,73){\makebox(0,0)[r]{$\vdots$}}
\put(-5,100){\makebox(0,0)[r]{$s$}}
\put(0,-5){\makebox(0,0)[t]{$0$}}
\put(30,-11){\makebox(0,0)[t]{$\ldots$}}
\put(60,-5){\makebox(0,0)[t]{$s\!-\!r$}}
\put(120,-11){\makebox(0,0)[t]{$\ldots$}}
\put(178,-5){\makebox(0,0)[t]{$N$}}
\put(185,0){\makebox(0,0)[l]{$1$}}
\put(185,23){\makebox(0,0)[l]{$\vdots$}}
\put(185,40){\makebox(0,0)[l]{$r$}}
\multiput(0,0)(0,20){6}{\line(1,0){180}}
\multiput(0,0)(20,0){10}{\line(0,1){100}}
\thicklines
\put(0,100){\line(1,-1){20}}
\put(20,80){\line(1,-1){20}}
\put(40,60){\line(1,-1){20}}
\put(60,40){\line(1,1){20}}
\put(80,60){\line(1,-1){20}}
\put(100,40){\line(1,1){20}}
\put(120,60){\line(1,-1){20}}
\put(140,40){\line(1,1){20}}
\put(160,60){\line(1,-1){20}}
\put(180,40){\line(1,1){20}}
\end{picture}
\end{center}
\caption{The $N$-step path $\sigma_h$ associated with the primary state $\ket h$ of the $(r,s)$ sector for (a)~$s\le r$  and (b)~$s>r$. The parity of $N$ is fixed by $N=|r-s|\ \mbox{mod $2$}$. Sectors of type (a) and (b) are related by the $\Z_2$ Kac table symmetry $(r,s)\equiv (L-r,L+1-s)$ under height reversal.}
\end{figure}

These RSOS paths were first introduced in the context of the Corner Transfer Matrices (CTMs) for the off-critical RSOS models but we argue in Section~4 that the same paths appear naturally at criticality in classifying the eigenstates of the critical RSOS double row transfer matrices.  Each path is associated with a state $|\sigma\rangle$ with configurational energy 
\begin{equation}
E(\sigma)=\half \sum_{j=1}^{N} j\,H(\sigma_{j-1},\sigma_j,\sigma_{j+1})
\label{energies}
\end{equation}
For the unitary minimal models ${\cal M}(L,L+1)$ the elementary local excitation energy is 
\begin{equation} 
H(\sigma_{j-1},\sigma_j,\sigma_{j+1})=\half |\sigma_{j-1}-\sigma_{j+1}|=
\begin{cases}
0,&\sigma_{j+1}=\sigma_{j-1}\\
1,&\sigma_{j+1}-\sigma_{j-1}=\pm 2
\end{cases}
\label{FlocalH}
\end{equation}
This description originates with the off-critical RSOS  
models in Regime~III with $q$ the elliptic modulus but here we apply it to the Regime III/IV critical RSOS models with $q$ the modular parameter. 
The vacuum or groundstate $\ket 0$  with zero energy is the path that alternates between heights 1 and 2.   In the $(r,s)$ sector, the primary or reference state $\ket h$ is the path $\sigma_h$ with lowest energy as shown in Figure~3.
The $(r,s)$ sector is associated with the Virasoro module $\V_h$ with $h=h_{r,s}$.
Specifically, the finitized Virasoro characters $\chi^{(N)}_{r,s}(q)=\chi^{(N)}_h(q)$ approach the corresponding Virasoro characters in the limit $N\to\infty$ with 
$N=|r-s|\ \mbox{mod $2$}$
\begin{equation}
\chi^{(N)}_h(q):=q^{-c/24+h}\sum_{\{\sigma\}} q^{E(\sigma)-E(\sigma_h)}
\to \chi_h(q),\qquad N\to\infty
\label{finchar}
\end{equation}


Let us just consider the RSOS models related to unitary minimal models in the sector $h=h_{r,s}$. 
Then at each position $j$ along an $A_L$  path $\sigma$ there is either a corner with energy 0 or  a straight segment with energy 1 and these two possibilities are mutually exclusive.
We regard an elementary excitation as the action of a fermionic operator $b_{\mp j/2}=b^h_{\mp j/2}$ 
that annihilates (creates) a corner at position $j$ of $\sigma$ and creates (annihilates) a straight segment at the same position of $\sigma$.
The fermionic behaviour (Pauli exclusion principle) is suggested by the fact that 
at position $j$ only one straight segment or corner can exist
\begin{equation} \label{action}
\setlength{\unitlength}{0.21mm}
\begin{array}{lll} 
\hspace{14pt} \begin{tabular}{ccc}
\begin{picture}(30,30)
\put(-20,-20){\line(1,1){20}}
\put(0,0){\circle*{3}}
\put(-3,-20){$\scriptstyle j$}
\put(0,0){\line(1,-1){20}}
\end{picture} & $\displaystyle {b_{-\frac{j}{2}} \atop \longmapsto }$ &
\hspace{14pt} \begin{picture}(30,30)
\put(-20,-20){\line(1,1){20}}
\put(0,0){\circle*{3}}
\put(-3,-20){$\scriptstyle j$}
\put(0,0){\line(1,1){20}}
\end{picture}  \end{tabular} & \rule{40pt}{0pt} 
\begin{tabular}{ccc}
\begin{picture}(30,30)
\put(-20,-20){\line(1,1){20}}
\put(0,0){\circle*{3}} \put(-3,-20){$\scriptstyle j$}
\put(0,0){\line(1,1){20}}
\end{picture} & $\displaystyle {b_{\frac{j}{2}} \atop \longmapsto }$ &
\hspace{14pt} \begin{picture}(30,30)
\put(-20,-20){\line(1,1){20}}
\put(0,0){\circle*{3}} \put(-3,-20){$\scriptstyle j$}
\put(0,0){\line(1,-1){20}}
\end{picture}  \end{tabular}  &\\[0pt]
\hspace{14pt}\begin{tabular}{ccc}
\begin{picture}(30,30)
\put(-20,20){\line(1,-1){20}}
\put(0,0){\circle*{3}} \put(-3,-20){$\scriptstyle j$}
\put(0,0){\line(1,1){20}}
\end{picture} & $\displaystyle {b_{-\frac{j}{2}} \atop \longmapsto }$ &
\hspace{14pt} \begin{picture}(30,30)
\put(-20,20){\line(1,-1){20}}
\put(0,0){\circle*{3}} \put(-3,-20){$\scriptstyle j$}
\put(0,0){\line(1,-1){20}}
\end{picture}
\end{tabular}  & \rule{40pt}{0pt} 
\begin{tabular}{ccc}
\begin{picture}(30,30)
\put(-20,20){\line(1,-1){20}}
\put(0,0){\circle*{3}} \put(-3,-20){$\scriptstyle j$}
\put(0,0){\line(1,-1){20}}
\end{picture} & $\displaystyle {b_{\frac{j}{2}} \atop \longmapsto }$ &
\hspace{14pt} \begin{picture}(30,30)
\put(-20,20){\line(1,-1){20}}
\put(0,0){\circle*{3}} \put(-3,-20){$\scriptstyle j$}
\put(0,0){\line(1,1){20}}
\end{picture}
\end{tabular} &
\hspace{40pt} \raisebox{18pt}{$b_{\pm\frac{j}{2}}^2 =0$}
\end{array}
\end{equation}
The associated energy  $H(\sigma_{j-1},\sigma_j,\sigma_{j+1})$ is the eigenvalue of the 
fermion number operator
\begin{equation}\label{benergy}
b_{-{j\over 2}}b_{{j\over 2}} |\sigma\rangle
=H(\sigma_{j-1},\sigma_j,\sigma_{j+1})|\sigma\rangle
\end{equation}
where in anticipation of unitarity we have set $b_{j/2}=b^\dagger_{-j/2}$.

The dual description originates with the RSOS models in the off-critical Regime~II but here we apply it to the critical Regime I/II RSOS models describing ${\mathbb Z}_{L-1}$ parafermions. In the dual description the roles of the corners and straight segments are interchanged so that
\begin{equation}
H'(\sigma_{j-1},\sigma_j,\sigma_{j+1})=1-H(\sigma_{j-1},\sigma_j,\sigma_{j+1})=
\begin{cases}
0,&\sigma_{j+1}-\sigma_{j-1}=\pm 2\\
1,&\sigma_{j+1}=\sigma_{j-1}
\end{cases}\label{energydual}
\end{equation}
Apart from a shift in the zero of energy, this involves an overall change of sign in the energy $E(\sigma)$. The groundstate $\ket 0'$ with zero energy is now the saw-tooth path  (see Figure~5) $\sigma=\{1,2,\ldots,L-1,L,L-1,\ldots, 2,1,2,\ldots\}$. In this dual picture an elementary excitation annihilates a straight segment and creates a corner at position $j$ of the path. The associated energy $H'(\sigma_{j-1},\sigma_j,\sigma_{j+1})$ is the eigenvalue of the dual fermion number operator
\begin{equation}
b_{{j\over 2}}b_{-{j\over 2}}|\sigma\rangle=
H'(\sigma_{j-1},\sigma_j,\sigma_{j+1})|\sigma\rangle
\end{equation}
Comparison with (\ref{FlocalH}) and (\ref{energydual}) suggests that
\begin{equation}
b_{-{j\over 2}}b_{{j\over 2}}+b_{{j\over 2}}b_{-{j\over 2}}=1
\end{equation}
consistent with our fermionic interpretation. Guided by the Ising or free fermion $\M(3,4)$ case, we assume the additional fermionic relations $b_j b_k=-b_k b_j$, $j\ne -k$.

We now introduce a fermionic algebra generated by all the fermion operators
\begin{equation}
\F=\oplus_h \F_h,\qquad \F_h=\langle b^h_j,\ j\in\Z/2, j\ne 0\rangle
\end{equation}
There is an independent copy $\F_h$ of the fermion algebra in each sector $h$ with
\begin{equation}
b^h_j\,b_k^{h'}=b^h_j\,b^h_k\,\delta_{h,h'}
\end{equation} 
We usually work in a fixed sector and suppress the index $h$.
The generators in each sector satisfy the usual Canonical Anticommutator Relations (CAR) for fermions as well as the involutive property for real fermions
\begin{equation}
\{b_j, b_k\}=\delta_{j,-k}, \qquad b_{-j}=b_{j}^{\dagger}=b_j^T
\label{CAR}
\end{equation}
The fermion number operators $b_{-j}b_j$ are mutually commuting projectors
\begin{equation}\label{fermionnumber}
(b_{-j}b_j)^2=b_{-j}b_j; \qquad [b_{-j}b_j,b_{-k}b_k]=0, \quad  j, \; k \neq 0;
\qquad [b_{-j}b_j,b_k]=0,\quad  j\neq \pm k
\end{equation}
On the lattice this algebra is {\it finitized} by setting
\begin{equation}
b_{j\over 2}=0,\qquad |j|>N
\end{equation}
This yields a closed finite-dimensional fermionic algebra
\begin{equation}
\F^{(N)}=\oplus_h \F_h^{(N)},\qquad\F_h^{(N)}
=\langle b^h_j, \   j=\mbox{\small $\pm {1\over 2},\pm 1,\ldots,\pm{N\over 2}$}\rangle
\end{equation}

A matrix representation of the fermionic algebra $\F_h^{(N)}$ is easily obtained in terms of $2^N\times 2^N$ real matrices with entries $0,\pm 1$.  The dimension of this algebra is $2^N$ since at each position there can be a straight segment or a corner (\ref{action}), that is, at each position a monomial in the basis can contain a $b_{-{j/2}}$ or the identity $1$. 

\setlength{\unitlength}{0.24mm}
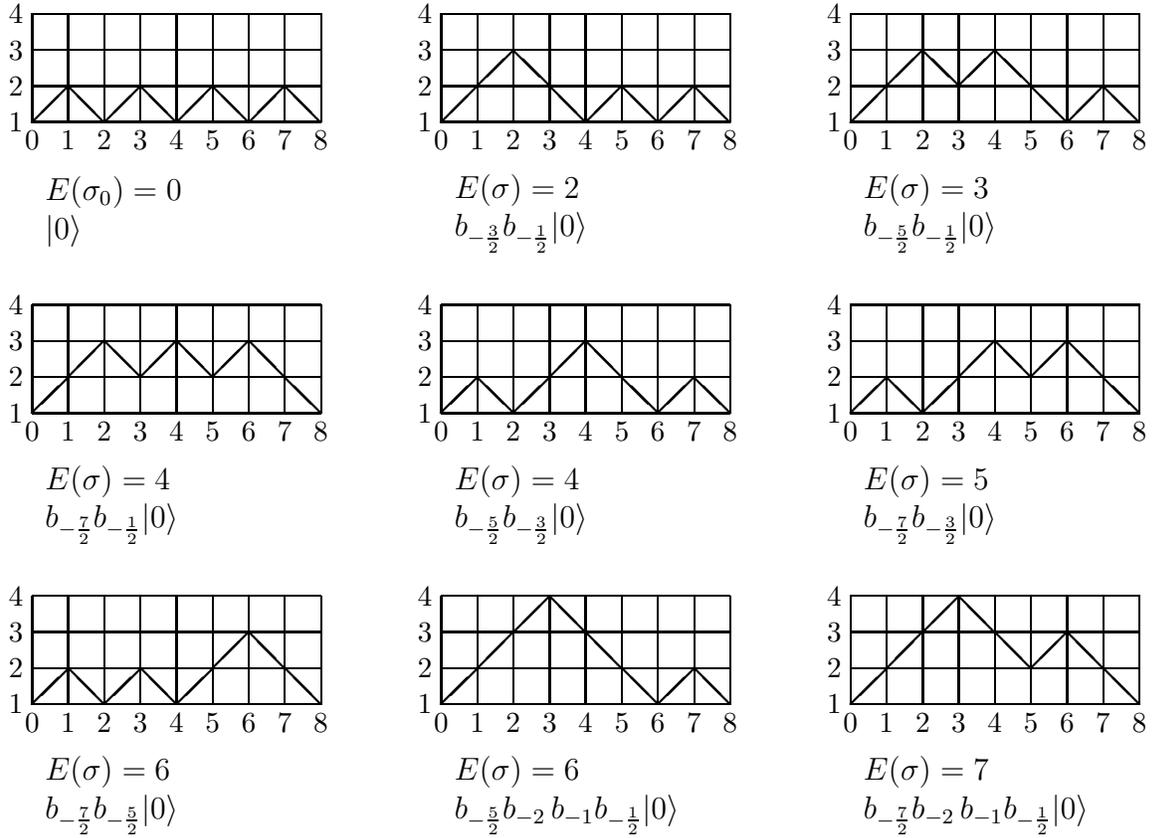
\begin{figure}[htb]
\begin{center}
\begin{tabular}{l@{\hspace{16mm}}l@{\hspace{16mm}}l}
\begin{picture}(160,80)
\lattices
\thicklines
\put(0,0){\line(1,1){20}}
\put(20,20){\line(1,-1){20}}
\put(40,0){\line(1,1){20}}
\put(60,20){\line(1,-1){20}}
\put(80,0){\line(1,1){20}}
\put(100,20){\line(1,-1){20}}
\put(120,0){\line(1,1){20}}
\put(140,20){\line(1,-1){20}}
\end{picture} &  
\begin{picture}(160,80)
\lattices
\thicklines
\put(0,0){\line(1,1){20}}
\put(20,20){\line(1,1){20}}
\put(40,40){\line(1,-1){20}}
\put(60,20){\line(1,-1){20}}
\put(80,0){\line(1,1){20}}
\put(100,20){\line(1,-1){20}}
\put(120,0){\line(1,1){20}}
\put(140,20){\line(1,-1){20}}
\end{picture} & 
\begin{picture}(160,80)
\lattices
\thicklines
\put(0,0){\line(1,1){20}}
\put(20,20){\line(1,1){20}}
\put(40,40){\line(1,-1){20}}
\put(60,20){\line(1,1){20}}
\put(80,40){\line(1,-1){20}}
\put(100,20){\line(1,-1){20}}
\put(120,0){\line(1,1){20}}
\put(140,20){\line(1,-1){20}}
\end{picture} \\[5mm]
$ \begin{array}{l} E(\sigma_0)=0 \\  \ket0 \end{array} $ &  
$\begin{array}{l} E(\sigma)=2 \\ b_{-{3\over2}} b_{-{1\over2}} \ket0 \end{array}$ & 
$\begin{array}{l} E(\sigma)=3 \\ b_{-{5\over2}} b_{-{1\over2}} \ket0 \end{array} $ \\[5mm]
\begin{picture}(160,80)
\lattices
\thicklines
\put(0,0){\line(1,1){20}}
\put(20,20){\line(1,1){20}}
\put(40,40){\line(1,-1){20}}
\put(60,20){\line(1,1){20}}
\put(80,40){\line(1,-1){20}}
\put(100,20){\line(1,1){20}}
\put(120,40){\line(1,-1){20}}
\put(140,20){\line(1,-1){20}}
\end{picture}  &
\begin{picture}(160,80)
\lattices
\thicklines
\put(0,0){\line(1,1){20}}
\put(20,20){\line(1,-1){20}}
\put(40,0){\line(1,1){20}}
\put(60,20){\line(1,1){20}}
\put(80,40){\line(1,-1){20}}
\put(100,20){\line(1,-1){20}}
\put(120,0){\line(1,1){20}}
\put(140,20){\line(1,-1){20}}
\end{picture} &
\begin{picture}(160,80)
\lattices
\thicklines
\put(0,0){\line(1,1){20}}
\put(20,20){\line(1,-1){20}}
\put(40,0){\line(1,1){20}}
\put(60,20){\line(1,1){20}}
\put(80,40){\line(1,-1){20}}
\put(100,20){\line(1,1){20}}
\put(120,40){\line(1,-1){20}}
\put(140,20){\line(1,-1){20}}
\end{picture}  \\[5mm]
$\begin{array}{l} E(\sigma)=4 \\ b_{-{7\over2}} b_{-{1\over2}} \ket0 \end{array} $ &
$\begin{array}{l} E(\sigma)=4 \\ b_{-{5\over2}} b_{-{3\over2}} \ket0 \end{array} $ &
$\begin{array}{l} E(\sigma)=5 \\ b_{-{7\over2}} b_{-{3\over2}} \ket0 \end{array} $ \\[5mm]
\begin{picture}(160,80)
\lattices
\thicklines
\put(0,0){\line(1,1){20}}
\put(20,20){\line(1,-1){20}}
\put(40,0){\line(1,1){20}}
\put(60,20){\line(1,-1){20}}
\put(80,0){\line(1,1){20}}
\put(100,20){\line(1,1){20}}
\put(120,40){\line(1,-1){20}}
\put(140,20){\line(1,-1){20}}
\end{picture}  &
\begin{picture}(160,80)
\lattices
\thicklines
\put(0,0){\line(1,1){20}}
\put(20,20){\line(1,1){20}}
\put(40,40){\line(1,1){20}}
\put(60,60){\line(1,-1){20}}
\put(80,40){\line(1,-1){20}}
\put(100,20){\line(1,-1){20}}
\put(120,0){\line(1,1){20}}
\put(140,20){\line(1,-1){20}}
\end{picture} &
\begin{picture}(160,80)
\lattices
\thicklines
\put(0,0){\line(1,1){20}}
\put(20,20){\line(1,1){20}}
\put(40,40){\line(1,1){20}}
\put(60,60){\line(1,-1){20}}
\put(80,40){\line(1,-1){20}}
\put(100,20){\line(1,1){20}}
\put(120,40){\line(1,-1){20}}
\put(140,20){\line(1,-1){20}}
\end{picture}  \\[5mm]
$\begin{array}{l} E(\sigma)=6 \\ b_{-{7\over2}} b_{-{5\over2}} \ket0 \end{array} $ &
$\begin{array}{l} E(\sigma)=6 \\ b_{-{5\over2}} b_{-2}\,b_{-1} b_{-{1\over2}} \ket0 \end{array} $ &
$\begin{array}{l} E(\sigma)=7 \\ b_{-{7\over2}} b_{-2}\,b_{-1} b_{-{1\over2}} \ket0 \end{array} $
\end{tabular}
\end{center}
\caption{The first few RSOS configurational paths for the vacuum sector $\mathcal{V}_0^{(8)}$ of 
$\mathcal{M}(4,5)$ with energy given by (\ref{FlocalH}). These states are also listed in Figure~7. 
The rule (\ref{action}) associates physical fermionic states with paths.}
\end{figure}

Each allowed path is naturally associated with a fermionic monomial (see Figures~4 and 5)
\begin{equation}
\label{monomial}
\{\sigma\}_{j_1,j_2,j_3,\ldots,j_n} \leftrightarrow 
b_{\mp{j_n\over 2}} \ldots b_{\mp{j_3\over 2}}b_{\mp{j_2\over 2}} b_{\mp{j_1\over 2}} \in 
\F_h, \qquad j_n>\ldots>j_3>j_2>j_1>0
\end{equation}
(anticommutation of the fermion operators has been used to bring them into canonical order) according to the following rule 
as in (\ref{action}):
\begin{quote}
The path $\sigma_{h}$ corresponding to the primary state $\ket h$ is associated to the identity in the fermionic algebra $\F_h$.

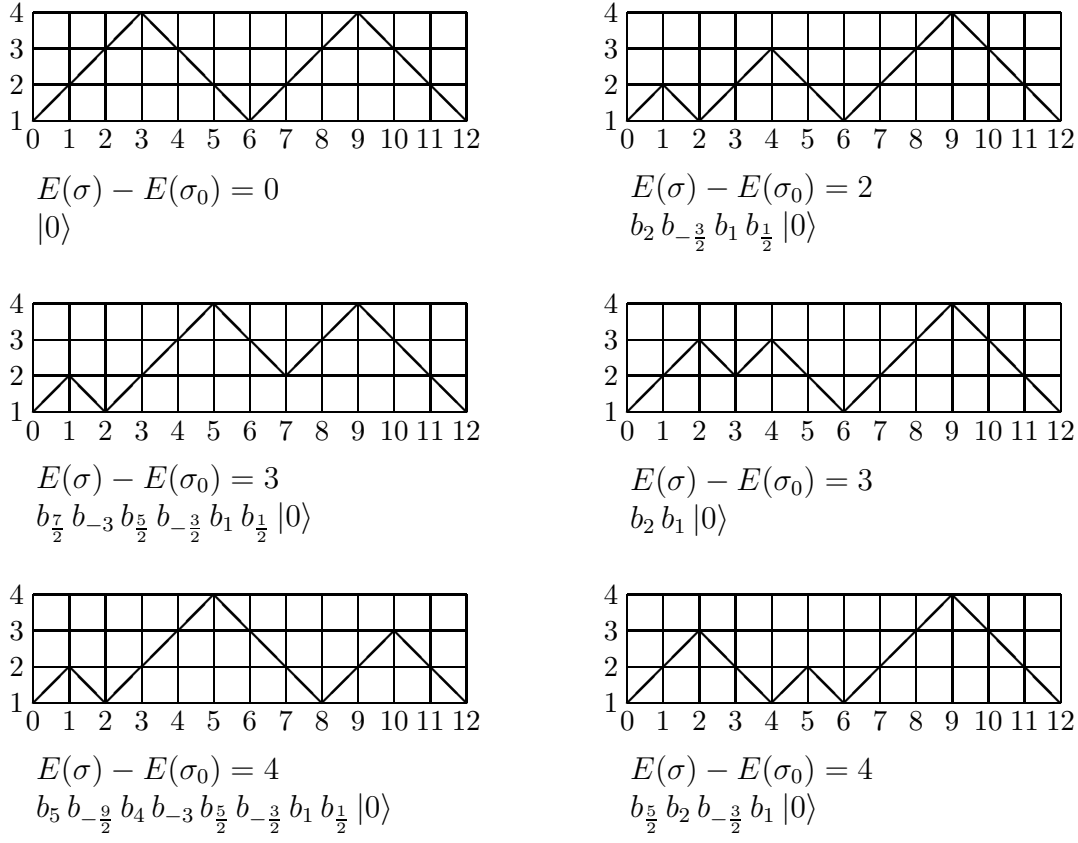
\begin{figure}[thb]
\begin{center}
\begin{tabular}{l@{\hspace{20mm}}l}
\setlength{\unitlength}{0.24mm}
\begin{picture}(240,80)
\latticel
\thicklines
\put(0,0){\line(1,1){20}}
\put(20,20){\line(1,1){20}}
\put(40,40){\line(1,1){20}}
\put(60,60){\line(1,-1){20}}
\put(80,40){\line(1,-1){20}}
\put(100,20){\line(1,-1){20}}
\put(120,0){\line(1,1){20}}
\put(140,20){\line(1,1){20}}
\put(160,40){\line(1,1){20}}
\put(180,60){\line(1,-1){20}}
\put(200,40){\line(1,-1){20}}
\put(220,20){\line(1,-1){20}}
\end{picture} &  
\setlength{\unitlength}{0.24mm}
\begin{picture}(240,80)
\latticel
\thicklines
\put(0,0){\line(1,1){20}}
\put(20,20){\line(1,-1){20}}
\put(40,0){\line(1,1){20}}
\put(60,20){\line(1,1){20}}
\put(80,40){\line(1,-1){20}}
\put(100,20){\line(1,-1){20}}
\put(120,0){\line(1,1){20}}
\put(140,20){\line(1,1){20}}
\put(160,40){\line(1,1){20}}
\put(180,60){\line(1,-1){20}}
\put(200,40){\line(1,-1){20}}
\put(220,20){\line(1,-1){20}}
\end{picture} \\[5mm]
$ \begin{array}{l} E(\sigma)-E(\sigma_0)=0 \\  \ket0 \end{array} $ &  
$\begin{array}{l} E(\sigma)-E(\sigma_0)=2 \\ b_2\,b_{-{3\over2}}\,b_1\,b_{1\over2}\,\ket0 \end{array}$ \\[5mm]
\setlength{\unitlength}{0.24mm}
\begin{picture}(240,80)
\latticel
\thicklines
\put(0,0){\line(1,1){20}}
\put(20,20){\line(1,-1){20}}
\put(40,0){\line(1,1){20}}
\put(60,20){\line(1,1){20}}
\put(80,40){\line(1,1){20}}
\put(100,60){\line(1,-1){20}}
\put(120,40){\line(1,-1){20}}
\put(140,20){\line(1,1){20}}
\put(160,40){\line(1,1){20}}
\put(180,60){\line(1,-1){20}}
\put(200,40){\line(1,-1){20}}
\put(220,20){\line(1,-1){20}}
\end{picture} &
\setlength{\unitlength}{0.24mm}
\begin{picture}(240,80)
\latticel
\thicklines
\put(0,0){\line(1,1){20}}
\put(20,20){\line(1,1){20}}
\put(40,40){\line(1,-1){20}}
\put(60,20){\line(1,1){20}}
\put(80,40){\line(1,-1){20}}
\put(100,20){\line(1,-1){20}}
\put(120,0){\line(1,1){20}}
\put(140,20){\line(1,1){20}}
\put(160,40){\line(1,1){20}}
\put(180,60){\line(1,-1){20}}
\put(200,40){\line(1,-1){20}}
\put(220,20){\line(1,-1){20}}
\end{picture}    \\[5mm]
$\begin{array}{l} E(\sigma)-E(\sigma_0)=3 \\ b_{7\over2}\,b_{-3}\,b_{5\over2}\,b_{-{3\over2}}\,b_1\,
     b_{1\over2}\,\ket0 \end{array} $ &
$\begin{array}{l} E(\sigma)-E(\sigma_0)=3 \\ b_2\, b_1\, \ket0 \end{array} $ \\[5mm]
\setlength{\unitlength}{0.24mm}
\begin{picture}(240,80)
\latticel
\thicklines
\put(0,0){\line(1,1){20}}
\put(20,20){\line(1,-1){20}}
\put(40,0){\line(1,1){20}}
\put(60,20){\line(1,1){20}}
\put(80,40){\line(1,1){20}}
\put(100,60){\line(1,-1){20}}
\put(120,40){\line(1,-1){20}}
\put(140,20){\line(1,-1){20}}
\put(160,0){\line(1,1){20}}
\put(180,20){\line(1,1){20}}
\put(200,40){\line(1,-1){20}}
\put(220,20){\line(1,-1){20}}
\end{picture} &
\setlength{\unitlength}{0.24mm}
\begin{picture}(240,80)
\latticel
\thicklines
\put(0,0){\line(1,1){20}}
\put(20,20){\line(1,1){20}}
\put(40,40){\line(1,-1){20}}
\put(60,20){\line(1,-1){20}}
\put(80,0){\line(1,1){20}}
\put(100,20){\line(1,-1){20}}
\put(120,0){\line(1,1){20}}
\put(140,20){\line(1,1){20}}
\put(160,40){\line(1,1){20}}
\put(180,60){\line(1,-1){20}}
\put(200,40){\line(1,-1){20}}
\put(220,20){\line(1,-1){20}}
\end{picture}    \\[5mm]
$\begin{array}{l} E(\sigma)-E(\sigma_0)=4 \\ 
       b_{5}\,b_{-{9\over2}}\,b_{4}\,b_{-3}\,b_{5\over2}\,b_{-{3\over2}}\,b_{1}\,b_{1\over2} \, \ket0 \end{array} $ &
$\begin{array}{l} E(\sigma)-E(\sigma_0)=4 \\ b_{5\over2}\,b_2\,b_{-{3\over2}}\,b_1\, \ket0 \end{array} $ 
\end{tabular}
\end{center}
\caption{\label{Potts_paths}The first few RSOS configurational paths in the parafermionic ${\Bbb Z}_3$ vacuum sector $\mathcal{V}_0 \oplus \mathcal{V}_3$
of the 3-state Potts model with energy given by (\ref{energydual}). 
The energy of the vacuum path (top-left corner) $\sigma_0=\{1,2,3,4,3,2,1,2,3,4,3,2,1\}$ has been subtracted. We used the usual association rule (\ref{action}) and  ordered the $b$ operators as they are obtained from the paths. The canonical order can be obtained by the anticommutation rules.}
\end{figure}

At each position $j$ where a corner of the primary state path is changed to a straight line we put a creation 
operator $b_{-\frac{j}{2}}$ and at each position $j$ where a straight line of the primary state path 
is changed to a corner we put an annihilation operator $b_{\frac{j}{2}}.$

No operators are inserted when a corner or straight line is unchanged.
\end{quote}
We call such a monomial a {\it physical} monomial. 
Among other constraints, the path restrictions imply
\begin{equation}
b_{-j}b_{-j-1/2}b_{-j-1}\ldots b_{-j-(L-2)/2}=\{\mbox{unphysical}\}
\end{equation}
so that such combinations, which go beyond the heights of $A_L$ and do not correspond to allowed paths, should never enter in the expressions of physical quantities and must be projected out. 

Note that the product of two physical monomials need not be a physical monomial. As an example, consider the two physical monomials and associated $A_4$ paths
\begin{eqnarray}
b_{-{5\over 2}}b_{-2} b_{-1}b_{-{1\over 2}} & \leftrightarrow & \{ 1,2,3,4,3,2,1,2,1 \}\nonumber \\
b_{-{7\over 2}}b_{-{3\over 2}}  & \leftrightarrow & \{ 1,2,1,2,3,2,3,2,1 \} \nonumber
\end{eqnarray}
We see that the product does not correspond to any $A_4$ path
\begin{equation}
(b_{-{5\over 2}}b_{-2} b_{-1}b_{-{1\over 2}}) (b_{-{7\over 2}}b_{-{3\over 2}}) 
= b_{-{7\over 2}} b_{-{5\over 2}} b_{-2} b_{-{3\over 2}} b_{-1} b_{-{1\over 2}}
\end{equation} 

\subsection{Physical fermionic states, finite Virasoro modules and projectors} 
\label{ferm.bas}

The identification (\ref{monomial}) between physical fermionic monomials and paths can be used to define physical fermionic states
\begin{equation}
\ket\sigma=b_{\mp{j_n\over 2}} \ldots b_{\mp{j_3\over 2}}b_{\mp{j_2\over 2}} b_{\mp{j_1\over 2}} \ket h, \qquad j_n>\ldots>j_3>j_2>j_1>0
\label{fermionstates}
\end{equation}
In the alternative view of infinite paths, the paths associated with these states must agree with the primary state $\ket h$ for all $j>N$ for some finite $N$. 
Normalizing the primary states so that $\langle h|h\rangle=1$, it is easy to verify that the states (\ref{fermionstates}) are orthonormal using the fermion algebra. 
Some physical fermionic states of $\mathcal{M}(4,5)$ and the 3-state Potts model are shown in Figures~4 and 5. 
A consequence of our definition of the primary state $\ket h$ is that it is annihilated by certain modes
\begin{equation}
\begin{array}{l@{~~}l}
 b_{\frac{j}{2}} |h\rangle =0  & \mbox{ if } \{ {\sigma_h} \}_j \mbox{ is a corner}  \\[2mm] 
 b_{-\frac{j}{2}} |h\rangle =0  & \mbox{ if } \{ {\sigma_h} \}_j \mbox{ is a straight line}  
\end{array}
\end{equation}
We identify the linear span of physical fermionic states with the Virasoro module $\V_h$. The set of states (\ref{fermionstates}) thus forms a fermionic basis $\B_h$ for $\V_h$. 
The Hilbert space of physical states decomposes into a direct sum of modules
\begin{equation}
\H=\oplus_h \V_h
\end{equation}
with fermionic basis $\B=\cup_h \B_h$. 
Again this space is graded according to the level
\begin{equation}
\V_h=\oplus_\ell \V_{h,\ell},\qquad \ell=\sum_{k=1}^n {j_k\over 2} - E(\sigma_h) =E(\sigma)-E(\sigma_h) 
\end{equation}
A generic Virasoro module of fermionic states in the $h=0$ vacuum sector is shown in Figure~6.

\setlength{\unitlength}{1pt}
\begin{figure}[p]
\hspace{.5in}
\begin{picture}(320,240)
\put(0,220){$\phantom{2}1$}\put(40,220){$\ket 0$}
\put(0,200){$\phantom{2}0$}\put(40,200){--}
\put(0,180){$\phantom{2}q^2$}\put(40,180){$b_{-{3\over 2}}b_{-{1\over 2}}\ket 0$}
\put(0,160){$\phantom{2}q^3$}
\put(40,160){$b_{-{5\over 2}}b_{-{1\over 2}}\ket 0$}
\put(0,140){$2q^4$}
\put(40,140){$b_{-{7\over 2}}b_{-{1\over 2}}\ket 0$}
\put(120,140){$b_{-{5\over 2}}b_{-{3\over 2}}\ket 0$}
\put(0,120){$2q^5$}
\put(40,120){$b_{-{9\over 2}}b_{-{1\over 2}}\ket 0$}
\put(120,120){$b_{-{7\over 2}}b_{-{3\over 2}}\ket 0$}
\put(0,100){$4q^6$}
\put(40,100){$b_{-{11\over 2}}b_{-{1\over 2}}\ket 0$}
\put(120,100){$b_{-{9\over 2}}b_{-{3\over 2}}\ket 0$}
\put(200,100){$b_{-{7\over 2}}b_{-{5\over 2}}\ket 0$}
\put(280,100){$b_{-{5\over 2}}b_{-2}b_{-1}b_{-{1\over 2}}\ket 0$}
\put(0,80){$4q^7$}
\put(40,80){$b_{-{13\over 2}}b_{-{1\over 2}}\ket 0$}
\put(120,80){$b_{-{11\over 2}}b_{-{3\over 2}}\ket 0$}
\put(200,80){$b_{-{9\over 2}}b_{-{5\over 2}}\ket 0$}
\put(280,80){$b_{-{7\over 2}}b_{-2}b_{-1}b_{-{1\over 2}}\ket 0$}
\put(0,60){$7q^8$}
\put(40,60){$b_{-{15\over 2}}b_{-{1\over 2}}\ket 0$}
\put(120,60){$b_{-{13\over 2}}b_{-{3\over 2}}\ket 0$}
\put(200,60){$b_{-{11\over 2}}b_{-{5\over 2}}\ket 0$}
\put(280,60){$b_{-{9\over 2}}b_{-{7\over 2}}\ket 0$}
\put(60,40){$b_{-{7\over 2}}b_{-{5\over 2}}b_{-{3\over 2}}b_{-{1\over 2}}\ket 0$}
\put(160,40){$b_{-{9\over 2}}b_{-2}b_{-1}b_{-{1\over 2}}\ket 0$}
\put(260,40){$b_{-{7\over 2}}b_{-3}b_{-1}b_{-{1\over 2}}\ket 0$}
\put(0,20){$8q^9$}
\put(40,20){$\ldots$}
\put(120,20){$\ldots$}
\put(200,20){$\ldots$}
\put(280,20){$\ldots$}
\put(0,0){$12q^{10}$}
\put(40,0){$\ldots$}
\put(120,0){$\ldots$}
\put(200,0){$\ldots$}
\put(280,0){$\ldots$}
\end{picture}
\caption{Generic Virasoro module $\V_0$ of orthonormal fermionic states in the  $h=0$ vacuum sector. The Virasoro character is $\chi_0(q)=\prod_{n=2}^\infty (1-q^n)^{-1}
=1+q^2+q^3+2q^4+2q^5+4q^6+4q^7+7q^8+8q^9+12q^{10}\cdots$. 
The first few states, such as, $b_{-{3\over 2}}b_{-{1\over 2}}\ket 0=\sqrt{2\over c}\,L_{-2}\ket 0$, $b_{-{5\over 2}}b_{-{1\over 2}}\ket 0={1\over \sqrt{2c}}\,L_{-3}\ket 0$, are easily related to Virasoro states. For finite $p$, unphysical (null) vectors appear. For example, for the Ising model $\M(3,4)$, 
$b_{-{5\over 2}}b_{-2}b_{-1}b_{-{1\over 2}}\ket 0$,
$b_{-{7\over 2}}b_{-2}b_{-1}b_{-{1\over 2}}\ket 0$,
$b_{-{9\over 2}}b_{-2}b_{-1}b_{-{1\over 2}}\ket 0$,
$b_{-{7\over 2}}b_{-3}b_{-1}b_{-{1\over 2}}\ket 0$ are null vectors. For $\M(4,5)$, the first null vector enters at level $12$.}
\end{figure}
\begin{figure}[p]
\hspace{.75in}
\begin{picture}(320,240)
\put(0,220){$\phantom{2}1$}\put(40,220){$\ket 0$}
\put(0,200){$\phantom{2}0$}\put(40,200){--}
\put(0,180){$\phantom{2}q^2$}\put(40,180){$b_{-{3\over 2}}b_{-{1\over 2}}\ket 0$}
\put(0,160){$\phantom{2}q^3$}
\put(40,160){$b_{-{5\over 2}}b_{-{1\over 2}}\ket 0$}
\put(0,140){$2q^4$}
\put(40,140){$b_{-{7\over 2}}b_{-{1\over 2}}\ket 0$}
\put(160,140){$b_{-{5\over 2}}b_{-{3\over 2}}\ket 0$}
\put(0,120){$q^5$}
\put(40,120){$b_{-{7\over 2}}b_{-{3\over 2}}\ket 0$}
\put(0,100){$2q^6$}
\put(40,100){$b_{-{7\over 2}}b_{-{5\over 2}}\ket 0$}
\put(160,100){$b_{-{5\over 2}}b_{-2}b_{-1}b_{-{1\over 2}}\ket 0$}
\put(0,80){$q^7$}
\put(40,80){$b_{-{7\over 2}}b_{-2}b_{-1}b_{-{1\over 2}}\ket 0$}
\put(0,60){$2q^8$}
\put(40,60){$b_{-{7\over 2}}b_{-{5\over 2}}b_{-{3\over 2}}b_{-{1\over 2}}\ket 0$}
\put(160,60){$b_{-{7\over 2}}b_{-3}b_{-1}b_{-{1\over 2}}\ket 0$}
\put(0,40){$q^9$}
\put(40,40){$b_{-{7\over 2}}b_{-3}b_{-2}b_{-{1\over 2}}\ket 0$}
\put(0,20){$q^{10}$}
\put(40,20){$b_{-{7\over 2}}b_{-3}b_{-2}b_{-{3\over 2}}\ket 0$}
\end{picture}
\caption{Finite Virasoro module $\V_0^{(8)}$ of fermionic states in the vacuum $h=0$ sector of $\M(4,5)$. For $N=8$ just 13 states survive after setting $b_j=0$ for $|j|\ge 4$. The finitized character is $\chi^{(8)}_{1,1}(q)
=1+q^2+q^3+2q^4+q^5+2q^6+q^7+2q^8+q^9+q^{10}$.}
\end{figure}

On the lattice, the {\it finitized} physical Hilbert space is
\begin{equation}
\H^{(N)}=\oplus_h \V_h^{(N)}
\end{equation}
where the {\it finite Virasoro module} $\V_h^{(N)}$ 
is the linear span of physical fermion states associated with $N$-step paths in the $(r,s)$ sector. 
The number of independent physical fermion states forming the basis $\B_h^{(N)}$ of $\V_h^{(N)}$ is thus equal to the number of paths from height $s$ to height $r$ in $N$ steps so that 
\begin{equation} \label{dimensionality}
\mbox{dim}\,\V_h^{(N)}=|\B_h^{(N)}|=[A^N]_{r,s}
\end{equation}
where $A$ is the adjacency matrix of $A_L$. The finite Virasoro module $\V_0^{(8)}$ of the $h=0$ vacuum sector of $\M(4,5)$ is shown in Figure~7.

The projectors onto the physical states in $\H$ and $\V_h$ are given by complete sums over the fermionic basis of states
\begin{equation}
\label{projector}
\P = \sum_{\ket\sigma\in {\mathcal B}} |\sigma\rangle\langle\sigma|,\qquad
\P_h = \sum_{\ket\sigma\in {\mathcal B_h}} |\sigma\rangle\langle\sigma|
\end{equation}
Consequently a physical operator $\phi$ on $\H$ is determined by its matrix elements  
between physical states
\begin{equation}
\phi=\P\,\phi\,\P=\sum_{\ket\sigma,\ket{\sigma'}\in {\mathcal B}}
\langle\sigma|\phi|\sigma'\rangle\; |\sigma\rangle\langle\sigma'|
\label{matrixelements}
\end{equation}
The elementary projectors are easily written in terms of fermions. For example, if $h=h_{3,1}$,
\begin{equation}
\ket 0\bra 0=\prod_{j=1}^\infty (1-b^0_{-{j\over 2}} b^0_{j\over 2})\in\F,\qquad
\ket h\bra h=(b^h_{-{1\over 2}} b^h_{1\over 2})(b^h_{-1} b^h_1)
\prod_{j=3}^\infty (1-b^h_{-{j\over 2}} b^h_{j\over 2})\in\F
\label{elemprojector}
\end{equation}
It follows that the operator $\phi$ can  be written entirely in terms of fermions once the matrix elements are known relative to a given fermionic basis. 
The algebra of physical observables is thus the subalgebra of the fermionic algebra $\F$ obtained by projection
\begin{equation}
{\cal A}=\P\F\P
\end{equation}
Since the structure of this algebra is complicated, however, it is often easier to work in the larger unphysical fermion algebra $\F$ and then project.

\subsection{Hamiltonian and characters}

The Hamiltonian of the fermionic system associated with the unitary minimal models in the $(r,s)$ sector is 
\begin{equation}
L_0= \P_h\big(\sum_{j=1}^\infty {j\over 2}\, b_{-{j\over 2}} b_{j\over 2} -E(\sigma_h)+ h\big)\P_h
\label{Hamiltonian}
\end{equation}
where $\P_h$ is the projector onto physical states.
We assert that each physical fermion state (\ref{fermionstates}) is an eigenstate of the Hamiltonian (\ref{Hamiltonian}) with eigenenergy given precisely by (\ref{energies}).
Indeed, using the fermionic algebra we find
\begin{equation}
L_0\ket\sigma=L_0\, b_{-{j_1\over 2}}b_{-{j_2\over 2}}\ldots b_{-{j_n\over 2}}\ket h
=\big(E(\sigma)-E(\sigma_h)+h\big)\;b_{-{j_1\over 2}}b_{-{j_2\over 2}}\ldots b_{-{j_n\over 2}}\ket h
\end{equation}
where
\begin{equation}
E(\sigma)=\half \sum_{j=1}^{N} j\,H(\sigma_{j-1},\sigma_j,\sigma_{j+1})
=\sum_{k=1}^n {j_k\over 2}
\end{equation}

The Virasoro character which is the generating function for the spectrum of the Virasoro module is now
\begin{equation}
\chi_h(q)= q^{-c/24}\, \Tr_{\V_h} q^{L_0} 
\end{equation}
The finitized characters, given by the finitization $b_{j\over 2}=0$ for $|j|>N$, are
\begin{equation}
\chi^{(N)}_h(q)=q^{-c/24}\, \Tr_{\V_h^{(N)}}\,q^{L_0}
\end{equation}
These are the generating functions for the spectra of the finite Virasoro modules.

\section{Bijections of Paths and Strings}
\setcounter{equation}{0}
\label{section:bijection}

In this section we discuss the bijection between the fermionic paths of Section~3 and eigenstates of the critical RSOS double row transfer matrices. More specifically, we exhibit an energy preserving bijection between the RSOS paths and patterns of zeros classifying the eigenvalues of the double row transfer matrices. 
We use the work of Warnaar~\cite{Warnaar} to further establish a bijection with fermionic particles. 
Although they are equi-numerous, the fermionic RSOS paths that label the eigenstates of the transfer matrices are not the same as the RSOS paths on which the transfer matrices act. Theses states can only be related by a complicated orthogonal transformation. 
It also needs to be emphasized that our bijection is between the fermionic RSOS paths and the patterns of zeros of the eigenvalues of the double row transfer matrices $\D(u)$. Within the framework of the usual Bethe ansatz, a separate bijection exists~\cite{DasmaFoda} between the RSOS paths and the rigged configurations related to the patterns of zeros of Baxter's auxiliary matrix $\Q(u)$.

Although we assert that our bijections are general, we content ourselves here with illustrating the bijections in the vacuum $h=0$ sectors of the unitary minimal models. In these cases the classification of eigenstates of the double row transfer matrices and their accompanying patterns of zeros are completely known~\cite{OPW} in terms of $(m,n)$-systems~\cite{Berkovich,Melzer}. The full details of the bijections for the unitary minimal models in all $(r,s)$ sectors and with periodic boundary conditions will be given elsewhere~\cite{FP}. In the case of periodic boundaries, there are two copies of Virasoro and the  states are labelled by two RSOS paths.  In the plane of the complex spectral parameter $u$, these come from the distinct patterns of zeros in the upper and lower half-planes corresponding respectively to the left- and right-chiral halves. On the cylinder, the complex conjugation symmetry ensures that the patterns of zeros in the upper and lower half-planes are the same so there is only one RSOS path and one copy of the Virasoro algebra.

\subsection{Critical Ising model}

The central charge of the critical Ising $\M(3,4)$ model is $c=1/2$. Let us consider the vacuum sector with boundary condition $(r,s)=(1,1)$ and $h=0$. The excitation energies are given by the scaling limit of the eigenvalues $D(u)$ of the double-row transfer matrix $\D(u)$. The eigenvalues $D(u)$ are classified~\cite{OPW} according to their zeros in the complex $u$-plane falling in or on the boundary of the single analyticity strip
\begin{equation}
-\frac{\lambda}{2}<\Re(u)<\frac{3\lambda}{2}
\end{equation}
 where $\lambda=\pi/4$ is the crossing parameter. For finite $N$, the low lying excitations contain $1$-strings (zeros in the middle of the strip) and $2$-strings (pairs of zeros at the edges of the strip with the same imaginary part). By complex conjugation symmetry, the patterns of zeros in the upper and lower half planes are the same so we only consider the upper half-plane. 
It is found that if the number of $1$-strings in the upper half-plane is $m$ and the number of $2$-strings is $n$ then this string content satisfies the simple $(m,n)$-system
\begin{equation}
m+n={N\over 2}
\end{equation}
where $m$ and $N$ are even. 

The lowest  or groundstate energy $E=0$ occurs for $m=0$. 
Among excitations with given string content, the lowest energy $E=m^2/2$ occurs when all of the $1$-strings are further from the real axis than all of the $2$-strings.
Moreover, for given string content, it is found that all patterns of zeros obtained by permuting the vertical positions of the $1$- and $2$-strings actually occur among the eigenvalues. Each time the position of adjacent $1$- and $2$-strings is interchanged the energy is increased by one unit. It follows~\cite{OPW} that the spectrum generating function is given by the finitized character
\begin{equation}
\chi^{(N)}_0(q)=q^{-c/24} \sum_E q^E=q^{-c/24}\!\!\sum_{m=0,2,4,\ldots}\!\! q^{m^2/2}\, \gauss{N/2}{m}_q
\to q^{-c/24}\!\!\sum_{m=0,2,4,\ldots}\!\!{q^{m^2/2}\over (q)_m}=\chi_0(q)
\label{CIMfermionchar}
\end{equation}
where the $q$-binomials are
\begin{equation}
\gauss{m+n}{m}_q
=\begin{cases}
\dfrac{(q)_{m+n}}{(q)_{m}(q)_{n}}, &\quad m,n\ge 0 \\
0, &\quad \text{otherwise}\\
\end{cases}
\end{equation}
with $(q)_n$ given by (\ref{qfactorial}) and $(q)_{0}=1$. 

\subsection{Tricritical Ising model}

The central charge of the tricritical Ising $\M(4,5)$ model is $c=7/10$. Again let us consider the vacuum sector with boundary condition $(r,s)=(1,1)$ and $h=0$. 
In classifying~\cite{OPW} the eigenvalues $D(u)$ of the
double-row transfer matrix $\D(u)$ there are now two relevant analyticity strips in the complex $u$-plane  
\begin{equation}
(1)\quad -\frac{\lambda}{2}<\Re(u)<\frac{3\lambda}{2},\qquad 
(2)\quad 2\lambda<\Re(u)<4\lambda
\end{equation}
where the crossing parameter is $\lambda=\pi/5$. The excitations are again classified by the string content in the upper half $u$-plane
\begin{equation}
\begin{split}
m_i&=\text{\{number of $1$-strings in strip $i=1,2$\}}\\
n_i&=\text{\{number of $2$-strings in strip $i=1,2$\}}
\end{split}
\label{stringdefs}
\end{equation}
where $m_1$, $m_2$ and $N$ are even and satisfy the $(\vm,\vn)$ system
\begin{equation}
m_1+n_1={N+m_2\over 2},\qquad
m_2+n_2={m_1\over 2}
\end{equation}

The groundstate energy $E=0$ now corresponds to $m_1=m_2=0$. 
For excitations with given string content $(\vm,\vn)$, the lowest energy $E=(m_1^2-m_1m_2+m_2^2)/2$ occurs when,  in both strips~1 and 2, all of the 1-strings are further out from the real axis than all of the 2-strings in the strip. Bringing the location of a 1-string, in either strip~1 or 2, closer to the real axis by interchanging its location with an adjacent 2-string increments 
the excitation energy $E$ by one unit. It thus follows~\cite{OPW} that the spectrum generating function is given by the finitized character
\begin{eqnarray}
\chi^{(N)}_0(q)&=&q^{-c/24}\!\!\sum_{m_1,m_2=0,2,4,\ldots}\!\!
q^{(m_1^2-m_1m_2+m_2^2)/2} \gauss{(N+m_2)/2}{m_1}_q \gauss{m_1/2}{m_2}_q\nonumber\\
&\to&q^{-c/24}\!\!\sum_{m_1,m_2=0,2,4,\ldots}\!\!
{q^{(m_1^2-m_1m_2+m_2^2)/2}\over (q)_{m_1}} \gauss{m_1/2}{m_2}_q =\chi_0(q)
\label{TIMfermionchar}
\end{eqnarray}

\subsection{Unitary minimal models}

The critical and tricritical Ising models are the first two in the $A_L$ series of unitary minimal models $\M(L,L+1)$. Again considering the $(r,s)=(1,1)$ vacuum sector with $h=0$, the eigenvalues of the double row transfer matrices can be classified~\cite{FP} by the number of $1$-strings $m_i$ and the number of $2$-strings $n_i$ in $L-2$ analyticity strips. These must satisfy the $(\vm,\vn)$ system~\cite{Berkovich,Melzer}
\begin{equation}
\vm + \vn = \frac{1}{2}(N\ve_{1}+A\vm)
\label{mn}
\end{equation}
where $A$ is the adjacency matrix of $A_{L-2}$, $C=2I-A$ is the Cartan matrix and $\ve_1=(1,0,\ldots,0)$. Here $\vm$, $\vn$ and $\ve_1$ are $L-2$ dimensional vectors and each $m_i$ and $N$ must be even.
The finitized characters, generalizing (\ref{CIMfermionchar}) and (\ref{TIMfermionchar}), are~\cite{Berkovich,Melzer}
\begin{eqnarray}
\chi^{(N)}_0(q)&=&q^{-c/24}\sum_{(\vm,\vn)} 
q^{{1\over 4}\vm C\vm} \prod_{i=1}^{L-2} \gauss{m_i+n_i}{m_i}_q \nonumber\\
&\to&q^{-c/24}\!\!\sum_{m_i=0,2,4,\ldots}\!\!
{q^{{1\over 4}\vm C\vm}\over (q)_{m_1}} \prod_{i=2}^{L-2} \gauss{m_i+n_i}{m_i}_q
=\chi_0(q)
\label{fermionchars}
\end{eqnarray}
These finite characters agree precisely with the finitized characters (\ref{finchar}) and they reproduce the correct counting of states with $q=1$. The limiting forms of these characters, however, differ fundamentally from (\ref{bosonchar}) because the coefficients in the $q$-series (\ref{CIMfermionchar}),  (\ref{TIMfermionchar}) and (\ref{fermionchars}) are manifestly nonnegative. For this reason these forms are called {\em fermionic}~\cite{McCoyEtAl} in contrast to the {\em bosonic} form (\ref{bosonchar}). 

\subsection{Energy-preserving bijection}

\setlength{\unitlength}{1pt}
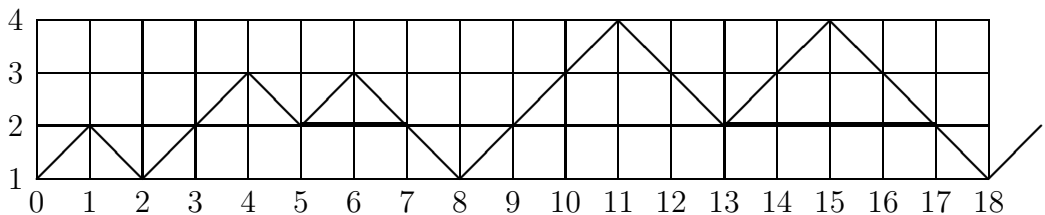
\begin{figure}[b]
\begin{center}
\begin{picture}(360,80)
\put(-5,0){\makebox(0,0)[r]{$1$}}
\put(-5,20){\makebox(0,0)[r]{$2$}}
\put(-5,40){\makebox(0,0)[r]{$3$}}
\put(-5,60){\makebox(0,0)[r]{$4$}}
\put(0,-5){\makebox(0,0)[t]{$0$}}
\put(20,-5){\makebox(0,0)[t]{$1$}}
\put(40,-5){\makebox(0,0)[t]{$2$}}
\put(60,-5){\makebox(0,0)[t]{$3$}}
\put(80,-5){\makebox(0,0)[t]{$4$}}
\put(100,-5){\makebox(0,0)[t]{$5$}}
\put(120,-5){\makebox(0,0)[t]{$6$}}
\put(140,-5){\makebox(0,0)[t]{$7$}}
\put(160,-5){\makebox(0,0)[t]{$8$}}
\put(180,-5){\makebox(0,0)[t]{$9$}}
\put(200,-5){\makebox(0,0)[t]{$10$}}
\put(220,-5){\makebox(0,0)[t]{$11$}}
\put(240,-5){\makebox(0,0)[t]{$12$}}
\put(260,-5){\makebox(0,0)[t]{$13$}}
\put(280,-5){\makebox(0,0)[t]{$14$}}
\put(300,-5){\makebox(0,0)[t]{$15$}}
\put(320,-5){\makebox(0,0)[t]{$16$}}
\put(340,-5){\makebox(0,0)[t]{$17$}}
\put(360,-5){\makebox(0,0)[t]{$18$}}
\multiput(0,0)(0,20){4}{\line(1,0){360}}
\multiput(0,0)(20,0){19}{\line(0,1){60}}
\thicklines
\put(100,21){\line(1,0){40}}
\put(260,21){\line(1,0){80}}
\put(0,0){\line(1,1){20}}
\put(20,20){\line(1,-1){20}}
\put(40,0){\line(1,1){20}}
\put(60,20){\line(1,1){20}}
\put(80,40){\line(1,-1){20}}
\put(100,20){\line(1,1){20}}
\put(120,40){\line(1,-1){20}}
\put(140,20){\line(1,-1){20}}
\put(160,0){\line(1,1){20}}
\put(180,20){\line(1,1){20}}
\put(200,40){\line(1,1){20}}
\put(220,60){\line(1,-1){20}}
\put(240,40){\line(1,-1){20}}
\put(260,20){\line(1,1){20}}
\put(280,40){\line(1,1){20}}
\put(300,60){\line(1,-1){20}}
\put(320,40){\line(1,-1){20}}
\put(340,20){\line(1,-1){20}}
\put(360,0){\line(1,1){20}}
\end{picture}
\end{center}
\caption{A typical $A_4$ RSOS path of $N=18$ steps decomposed into a series of overlapping peaks by drawing in a sequence of horizontal baselines following the prescription of Warnaar~\cite{Warnaar}.  This prescription reveals particles of type $1$, $2$, $1$, $3$, $2$ at positions $j=1,4,6,11,15$ respectively and a particle content $n_1=2, n_2=2, n_3=1$. 
A particle of type $i$ has a peak $i$ units above the baseline and a width (at the baseline) of $2i$.
\label{particles}}
\end{figure}
\setlength{\unitlength}{1pt}
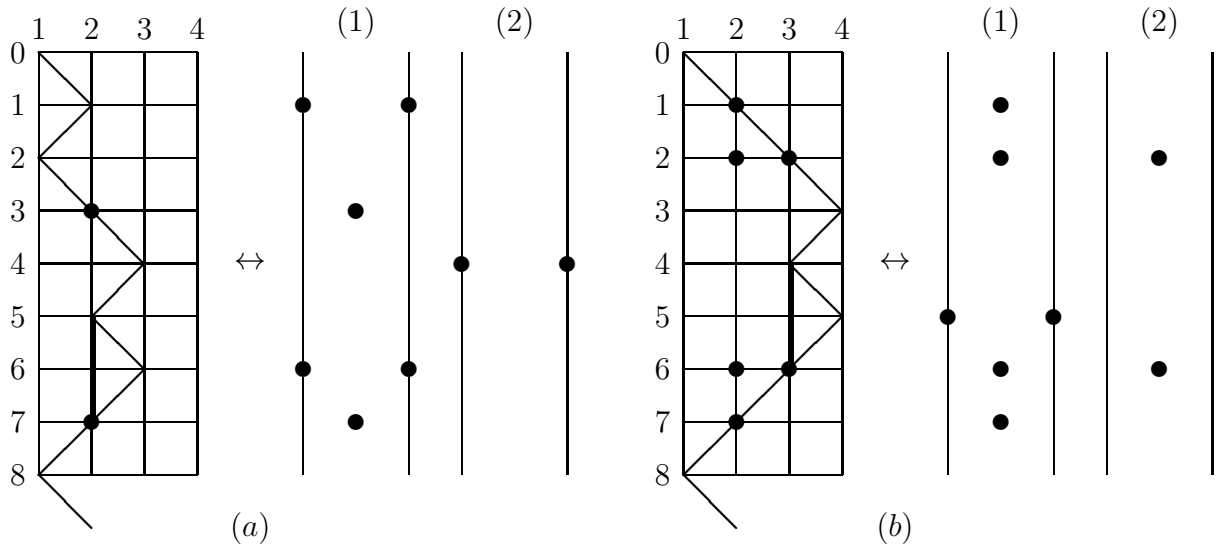
\begin{figure}[htb]
\begin{picture}(240,200)
\put(100,0){\makebox(0,0)[c]{$(a)$}}
\put(20,185){\makebox(0,0)[b]{$1$}}
\put(40,185){\makebox(0,0)[b]{$2$}}
\put(60,185){\makebox(0,0)[b]{$3$}}
\put(80,185){\makebox(0,0)[b]{$4$}}
\put(15,180){\makebox(0,0)[r]{$0$}}
\put(15,160){\makebox(0,0)[r]{$1$}}
\put(15,140){\makebox(0,0)[r]{$2$}}
\put(15,120){\makebox(0,0)[r]{$3$}}
\put(15,100){\makebox(0,0)[r]{$4$}}
\put(15,80){\makebox(0,0)[r]{$5$}}
\put(15,60){\makebox(0,0)[r]{$6$}}
\put(15,40){\makebox(0,0)[r]{$7$}}
\put(15,20){\makebox(0,0)[r]{$8$}}
\multiput(20,20)(0,20){9}{\line(1,0){60}}
\multiput(20,20)(20,0){4}{\line(0,1){160}}
\multiput(120,20)(40,0){2}{\line(0,1){160}}
\multiput(180,20)(40,0){2}{\line(0,1){160}}
\put(140,185){\makebox(0,0)[b]{$(1)$}}
\put(200,185){\makebox(0,0)[b]{$(2)$}}
\put(100,100){\makebox(0,0)[c]{$\leftrightarrow$}}
\multiput(120,60)(40,0){2}{\circle*{6}}
\multiput(120,160)(40,0){2}{\circle*{6}}
\multiput(180,100)(40,0){2}{\circle*{6}}
\put(40,40){\circle*{6}}
\put(40,120){\circle*{6}}
\put(140,40){\circle*{6}}
\put(140,120){\circle*{6}}
\thicklines
\put(40,40){\line(0,1){40}}
\put(41,40){\line(0,1){40}}
\put(20,20){\line(1,-1){20}}
\put(20,20){\line(1,1){20}}
\put(40,40){\line(1,1){20}}
\put(60,60){\line(-1,1){20}}
\put(40,80){\line(1,1){20}}
\put(60,100){\line(-1,1){20}}
\put(40,120){\line(-1,1){20}}
\put(20,140){\line(1,1){20}}
\put(40,160){\line(-1,1){20}}
\end{picture}
\begin{picture}(240,200)
\put(100,0){\makebox(0,0)[c]{$(b)$}}
\put(20,185){\makebox(0,0)[b]{$1$}}
\put(40,185){\makebox(0,0)[b]{$2$}}
\put(60,185){\makebox(0,0)[b]{$3$}}
\put(80,185){\makebox(0,0)[b]{$4$}}
\put(15,180){\makebox(0,0)[r]{$0$}}
\put(15,160){\makebox(0,0)[r]{$1$}}
\put(15,140){\makebox(0,0)[r]{$2$}}
\put(15,120){\makebox(0,0)[r]{$3$}}
\put(15,100){\makebox(0,0)[r]{$4$}}
\put(15,80){\makebox(0,0)[r]{$5$}}
\put(15,60){\makebox(0,0)[r]{$6$}}
\put(15,40){\makebox(0,0)[r]{$7$}}
\put(15,20){\makebox(0,0)[r]{$8$}}
\multiput(20,20)(0,20){9}{\line(1,0){60}}
\multiput(20,20)(20,0){4}{\line(0,1){160}}
\multiput(120,20)(40,0){2}{\line(0,1){160}}
\multiput(180,20)(40,0){2}{\line(0,1){160}}
\multiput(120,80)(40,0){2}{\circle*{6}}
\put(140,40){\circle*{6}}
\put(140,60){\circle*{6}}
\put(140,140){\circle*{6}}
\put(140,160){\circle*{6}}
\put(200,60){\circle*{6}}
\put(200,140){\circle*{6}}
\put(140,185){\makebox(0,0)[b]{$(1)$}}
\put(200,185){\makebox(0,0)[b]{$(2)$}}
\put(100,100){\makebox(0,0)[c]{$\leftrightarrow$}}
\put(40,40){\circle*{6}}
\put(40,60){\circle*{6}}
\put(60,60){\circle*{6}}
\put(40,140){\circle*{6}}
\put(60,140){\circle*{6}}
\put(40,160){\circle*{6}}
\thicklines
\put(20,20){\line(1,-1){20}}
\put(20,20){\line(1,1){20}}
\put(40,40){\line(1,1){20}}
\put(60,60){\line(1,1){20}}
\put(80,80){\line(-1,1){20}}
\put(60,100){\line(1,1){20}}
\put(80,120){\line(-1,1){20}}
\put(60,140){\line(-1,1){20}}
\put(40,160){\line(-1,1){20}}
\put(60,60){\line(0,1){40}}
\put(61,60){\line(0,1){40}}
\end{picture}
\caption{Bijection of tricritical Ising RSOS paths and strings. The RSOS paths (rotated  $90^\circ $  clockwise) are shown on the left and the strips (1) and (2) containing $1$- and $2$-strings in the upper-half complex $u$-plane are shown on the right. The string (particle) content is (a) $m_1=2$, $n_1=2$, $m_2=0$, $n_2=1$, $n_3=0$, (b) $m_1=4$, $n_1=1$, $m_2=2$, $n_2=0$, $n_3=1$.
\label{bijection}}
\end{figure}

Since the energy spectrum generating functions for the paths and strings coincide, it is natural to expect an energy-preserving bijection between the two descriptions of the states.
Given an RSOS path, Warnaar~\cite{Warnaar} has given a detailed prescription to extract the fermionic particle content. There are $L-1$ types of particles. A particle  of type $i$ at position $j$ corresponds to a peak in the RSOS path at height $i$ above the baseline as shown in Figure~\ref{particles}. For $1\le i\le L-2$, this correponds to a $2$-string in strip~$i$ at position $j$. Consequently, the number of particles $n_i$ of type $i$ coincides with the $n_i$ of the $(\vm,\vn)$ system for $i\le L-2$. Since the width of a particle of type $i$ is $2i$, the particle number $n_{L-1}$ is simply determined by the constraint
\begin{equation}
\sum_{i=1}^{L-1} 2i\, n_i=N
\end{equation}

The $m_i$ are interpreted as the number of dual-particles of type $i$ for $1\le i\le L-2$. In an RSOS path, a dual-particle of type $i$ at position $j$ corresponds to a domain wall (straight segment with $H(\sigma_{j-1},\sigma_j,\sigma_{j+1})=1$) at height $i$ above the baseline. In turn, this corresponds to $1$-strings at position $j$ in strips~1 through $i$. We remark that in the string description the absolute position of the strings in a strip is of no importance, only the relative ordering of $1$- and $2$-strings in each strip is pertinent. The rules of this bijection are illustrated in Figure~\ref{bijection} for two typical RSOS paths of the $A_4$ tricritical Ising model. In the case of the $\M(L,L+1)$ models the bijection works the same but there are $L-2$ such strips each containing $1$- and $2$-strings.

\section{Discussion}

In this paper we have introduced fermionic algebras into the study of $s\ell(2)$ minimal and parafermion conformal field theories by associating orthonormal fermionic states with the RSOS paths of the underlying lattice models. In this way we are able to build the Hilbert space of physical states of these models on the cylinder. The fermionic states have the advantage over Virasoro states in that they are automatically orthonormal. 

Our presentation here of the classification of eigenstates of the double row transfer matrices of the $A_L$ RSOS models and the details of the energy-preserving bijection with RSOS paths is incomplete. The task of extending this to all $(r,s)$ sectors of the unitary minimal models $\M(L,L+1)$ will be completed  elsewhere~\cite{FP}. Lastly, we have side-stepped the modifications required to treat the non-unitary models. We also hope to take this up in future work. 

In Part II of this series we will build, level by level, matrix representations of the Virasoro generators and the fields for the minimal models.
\goodbreak

\section*{Acknowledgements} 
\label{sec:Acknowledgements}
This research is supported by the Australian Research Council. We thank Omar Foda, Vladimir Rittenberg and Ole Warnaar for discussions and Jean-Bernard Zuber for comments on the manuscript.

\end{document}